\documentclass{PoS}

\usepackage{amsmath}
\usepackage{epsfig}
\usepackage{amssymb}
\usepackage{bbold}
\usepackage{braket}
\input{epsf}

\newcommand\as{\alpha_s}
\def\be{\begin{equation}}
\def\ee{\end{equation}}
\def\beq{\begin{equation}}
\def\eeq{\end{equation}}
\def\bea{\begin{eqnarray}}
\def\eea{\end{eqnarray}}
\def\ba{\begin{eqnarray}}
\def\ea{\end{eqnarray}}
\def\eeq{\end{equation}}
\def\beeq{\begin{eqnarray}}
\def\eeeq{\end{eqnarray}}
\def\beqa{\begin{eqnarray}}
\def\eeqa{\end{eqnarray}}

\newcommand\f{\frac}

\def\to{\rightarrow}
\def\nn{\nonumber}
\def\th{\hat{\tau}}

\title{Toward NNLL Resummation for Hadron Production in Hadronic Collisions \thanks{YITP-SB-15-42, LA-UR-15-27999}}

\ShortTitle{Toward NNLL Resummation for Hadron Production in Hadronic Collisions}

\author{Patriz Hinderer\\
        Institute for Theoretical Physics, University of T\"ubingen, Auf der Morgenstelle 14, 72076 T\"ubingen, Germany\\
        E-mail: \email{patriz.hinderer@uni-tuebingen.de}}

\author{\speaker{Felix Ringer}\\
        Theoretical Division, MS B283, Los Alamos National Laboratory, Los Alamos, NM 87545, USA\\
        E-mail: \email{f.ringer@lanl.gov}}

\author{George F. Sterman\\
        C.N.\ Yang Institute for Theoretical Physics, Stony Brook University, Stony Brook, New York 11794 -- 3840, USA \\
        E-mail: \email{george.sterman@stonybrook.edu}}

\author{Werner Vogelsang\\
        Institute for Theoretical Physics, University of T\"ubingen, Auf der Morgenstelle 14, 72076 T\"ubingen, Germany\\
        E-mail: \email{werner.vogelsang@uni-tuebingen.de}}

\abstract{We present results relevant for the extension of threshold resummation beyond the next-to-leading logarithmic (NLL) order for QCD hard-scattering processes. As an example, we consider di-hadron production $H_1 H_2\to h_1 h_2 X$, where the produced pair has a large invariant mass. Taking into account the non-trivial color structure of the partonic hard-scattering process, we determine the hard and soft matrices in color space. In our numerical studies we find a significant improvement compared to previous results at NLL accuracy. In particular, the scale dependence of the resummed cross section is greatly reduced. In addition, we comment on the extension of the techniques developed in this work to other observables relevant for hadronic collisions.}
          
\FullConference{QCD Evolution 2015\\
		          May 26-30, 2015\\
		          Jefferson Lab (JLAB), Newport News Virginia, USA}

\begin{document}

\section{Introduction}

The advances of precision measurements in hadron-hadron collisions carried out at experiments such as the LHC, RHIC and low energy fixed-target experiments, have lead to a growing interest in the resummation of threshold logarithms in the partonic hard-scattering cross sections. The partonic threshold is reached when the initial partons have just enough energy to produce the observed final state. In this work, we consider the hadronic di-hadron production cross section. In this case, the partonic threshold is reached when $\hat{s}=\hat{m}^2$, that is, $\hat\tau\equiv\hat{m}^2/\hat{s}=1$. Here, $\sqrt{\hat{s}}$
is the partonic center-of-mass system (c.m.s.) energy and $\hat{m}$ the invariant pair
mass of the two outgoing partons that fragment into the observed hadron pair. To some extend this process may be viewed as the ``full QCD'' extension of the Drell-Yan process, where the lepton pair in the final state is replaced by a hadron pair. The threshold logarithms in the perturbative series take the general form 
\be\label{series1}
\sum_{k=0}^\infty\sum_{\ell=1}^{2k} \as^k \,{\cal A}_{k,\ell} \, \left(\f{ \ln^{2k-\ell}
(1-\hat\tau)}{1-\hat\tau}\right)_+\,,
\ee
where $\as$ is the strong coupling constant, ${\cal A}_{k,\ell}$ are perturbatively calculable coefficients and the ``plus'' distribution will be defined below. The all-order set of logarithms with a fixed $\ell$ are often referred to as the $\ell$th {\it tower} of logarithms. Following the literature~\cite{dyresum,KS,BCMN}, threshold logarithms can be exponentiated, or ``resummed'', after taking an integral transform conjugate to the relevant kinematical variable (here $\hat\tau$). The accuracy of resummation is defined by counting the towers of logarithms that are fully under control. At next-to-leading logarithmic (NLL) accuracy three towers are under control and at next-to-next-to-leading logarithmic (NNLL) accuracy five towers are fully taken into account. Previously, Ref.~\cite{Almeida:2009jt} presented a NLL study for di-hadron production which forms the basis for our studies. In this work, we are going to extend the accuracy of resummation to the fourth tower, {\it i.e.} partial NNLL. This can only be achieved by taking into account the non-trivial color structure of the partonic QCD processes. For the first time, we derive all relevant ingredients for the extension of threshold resummation toward NNLL for a process where four colored partons are taking part in the scattering at leading-order. 

The partonic QCD scattering processes encountered in di-hadron production give the underlying structure for various other observables as well. Hence, di-hadron production is an ideal starting point for the study of QCD resummation beyond NLL and can serve as a template for reactions of more significant phenomenological interest. That said, di-hadron production is phenomenologically relevant in its own right as experimental data
as a function of the pair's mass are available from various fixed-target experiments~\cite{na24,e711,e706}, as well as from the ISR~\cite{ccor}. In addition, 
di-hadron cross sections are also accessible at the Relativistic Heavy Ion Collider (RHIC).

In Sec.~\ref{sec2} we present the basic formulas for the di-hadron cross section as a function of pair mass at fixed order in perturbation theory, and display the role of the threshold region. Section~\ref{sec3} presents several details of the NNLL threshold resummation for the cross section. In Sec.~\ref{sec4} we give phenomenological results, comparing the threshold resummed calculations at NLL and NNLL to some of 
the available experimental data. Finally, we summarize our results in Sec.~\ref{sec5}. The results reported in this paper are taken from~\cite{Hinderer:2014qta}, to which we refer the reader for further details.

\section{Hadron pair production near partonic threshold \label{sec2}}

\subsection{Perturbative cross section}

For sufficiently large invariant mass squared of the final state hadron pair $M^2$, the cross section for the process $H_1H_2 \to h_1h_2X$ can be written in a factorized form
\beeq
M^4\frac{d \sigma^{H_1 H_2\to h_1 h_2 X}}{dM^2 d\Delta \eta  d\bar{\eta} }
&=&\sum_{abcd}
\int_0^1 dx_a dx_b  
dz_c  dz_d  \, f_a^{H_1}
(x_a,\mu_F)f_b^{H_2}(x_b,\mu_F)\,  z_c D_c^{h_1}(z_c,\mu_F)
z_dD_d^{h_2}(z_d,\mu_F) \nn \\
&& \hspace{10mm}
\times  \, \omega_{ab\to cd} \left(\th, \Delta \eta, 
\hat{\eta}, \as(\mu_R), \frac{\mu_R}{\hat{m}},\frac{\mu_F}{\hat{m}}\right)\;.
\label{taufac} 
\eeeq
Here we defined the difference and average of the final state hadron c.m.s. rapidities $\eta_{1,2}$ as
\be\label{baretadef1}
\Delta \eta= \frac{1}{2}(\eta_1 -\eta_2)  \; , \quad \bar{\eta} =\frac{1}{2}(\eta_1 +\eta_2) \; ,
\ee
where $\Delta\eta$ is boost invariant and $\bar \eta$ is related to the average rapidity in the partonic c.m.s. $\hat\eta$ by
\beq
\hat{\eta}=\bar\eta-\frac{1}{2} \ln 
\left(\frac{x_a}{x_b}\right)  \; . 
\label{baretadef}
\eeq
The functions $f^{H_{1,2}}_{a,b}$ in Eq.~(\ref{taufac}) are the parton distribution functions
for partons $a,b$ in hadrons $H_{1,2}$ and $D_{c,d}^{h_{1,2}}$ are the 
fragmentation functions for partons $c,d$ fragmenting into the 
observed hadrons $h_{1,2}$. Furthermore, we define the variables
\be\label{eq:taus}
\hat\tau = \f{\hat m^2}{\hat s}\, ,\quad \tau'=\f{\hat m^2}{S} \, ,
\ee
where $S$ is the hadronic c.m.s.~energy. The variable $\hat\tau$ appears in the hard-scattering functions $\omega_{ab\to cd}$ in Eq.~(\ref{taufac}), whereas $\tau'$ will be used to below to define the Mellin transformation. The functions $\omega_{ab\to cd}$ may be computed in QCD perturbation theory
\beq\label{overall}
\omega_{ab\to cd} = \left( \frac{\alpha_s}{\pi}\right)^2 \left[ 
\omega_{ab\to cd}^{\mathrm{LO}} + \frac{\alpha_s}{\pi}\,
\omega_{ab\to cd}^{\mathrm{NLO}} + \left(\frac{\as}{\pi} \right)^2
\omega_{ab\to cd}^{\mathrm{NNLO}} + \ldots \right] \; .
\eeq
The limit $\hat{\tau} \to 1$ corresponds to the partonic threshold, where the hard-scattering uses all available energy to produce the pair. In general, as discussed in~\cite{Almeida:2009jt}, near partonic threshold, the $\omega_{ab\to cd}$ can be written as
\ba
\omega_{ab\to cd} \left(\th, \Delta \eta, 
\hat{\eta}, \alpha_s(\mu_R), \frac{\mu_R}{\hat{m}},\frac{\mu_F}{\hat{m}} \right)
& = & \delta \left(\hat{{\eta}} \right)\,
\omega^{{\mathrm{sing}}}_{ab\to cd}\left(\th, \Delta \eta, 
\alpha_s(\mu_R),\frac{\mu_R}{\hat{m}},\frac{\mu_F}{\hat{m}} \right)\nn \\[2mm]
&+&\omega_{ab\to cd}^{{\mathrm{reg}}} \left(\th, \Delta \eta, 
\hat{\eta}, \alpha_s(\mu_R), \frac{\mu_R}{\hat{m}},\frac{\mu_F}{\hat{m}} \right)
\; . \label{allord}
\ea
The function $\delta(\hat\eta)$ implies ``LO kinematics'' at threshold where $\bar\eta=\frac{1}{2} \ln(x_a/x_b)$ in Eq.~(\ref{baretadef}). All threshold logarithms in the perturbative series, see Eq.~(\ref{series1}), are contained in the functions $\omega^{{\mathrm{sing}}}_{ab\to cd}$. Threshold resummation addresses these logarithms to all orders in the strong coupling. All remaining contributions, which are subleading near threshold, are collected in the ``regular'' functions $\omega^{{\mathrm{reg}}}_{ab\to cd}$.

\subsection{Mellin and Fourier transforms}

In order to achieve the resummation of threshold logarithms, we take Fourier and Mellin integral transforms in the following way~\cite{Almeida:2009jt}. We separate the hard-scattering function and the PDFs from the fragmentation functions. We only take moments of the PDFs and the hard-scattering function, where the Mellin moments are taken with respect to $\tau^{\prime}$, defined in~(\ref{eq:taus}), and the Fourier transform is with respect to $\bar{\eta}$. One obtains
\beqa
\label{Omemom}
&&\hspace*{-.5cm}\sum_{ab}\int_{-\infty}^{\infty} 
d\bar\eta \, {\mathrm{e}}^{i \nu \bar{\eta}}\int_0^1 d\tau'\,
\left(\tau'\right)^{N-1} \int_0^1 d x_a \,d x_b \,
 f_a^{H_1} \left(x_a, \mu_F \right)\, f_b^{H_2}\left(x_b, \mu_F \right)\omega_{ab\to cd} \left( \hat{\tau},  \Delta \eta,\hat{{\eta}},
\as(\mu_R),\frac{\mu_R}{\hat{m}}, \frac{\mu_F}{\hat{m}} 
\right) \nn \\[2mm]
&&\hspace*{0mm}= \sum_{ab} \tilde{f}_a^{H_1}(N+1+i\nu/2,\mu_F)
\tilde{f}_b^{H_2}(N+1-i\nu/2,\mu_F)
\; \tilde{\omega}_{ab\to cd}
\left(N,\nu, \Delta \eta, \as(\mu_R),\frac{\mu_R}{\hat{m}}, \frac{\mu_F}{\hat{m}} 
\right)  \; , 
\eeqa
where $\tilde{f}_a^H(N,\mu_F)\equiv\int_0^1 x^{N-1}f_a^H(x,\mu_F)dx$, and 
\beq
\tilde{\omega}_{ab\to cd}
\left(N,\nu, \Delta \eta, \as(\mu_R),\frac{\mu_R}{\hat{m}}, \frac{\mu_F}{\hat{m}} 
\right)  \equiv \int_{-\infty}^{\infty}d\hat{\eta}\,
{\mathrm{e}}^{i \nu \hat{\eta}}\int_0^1 
d\hat\tau \,\hat\tau^{N-1} \, \omega_{ab\to cd}
\left(\hat{\tau}, \Delta \eta,\hat{\eta},\as(\mu_R),\frac{\mu_R}{\hat{m}}, \frac{\mu_F}{\hat{m}}  \right) \;.
\label{omegamom1}
\eeq
Due to the delta function $\delta(\hat\eta)$ in~(\ref{allord}), the $d\hat\eta$ integral becomes trivial near threshold. Threshold logarithms in $\omega_{ab\to cd}$, see Eq.~(\ref{series1}), will be transformed into logarithms of the Mellin variable $N$
\be
\as^k \left(\f{ \ln^{2k-\ell} (1-\hat\tau)}{1-\hat\tau}\right)_+ \to \as^k\ln^{2k-\ell+1}\bar N\, ,
\ee
where $\bar N=Ne^{\gamma_E}$ and the ``plus''-distributions are defined by
\beq
\int_{x_0}^1 f(x)\left( g(x)\right)_+ 
dx\equiv \int_{x_0}^1 \left (f(x) -f(1) \right) \, 
g(x) dx - f(1) \int_0^{x_0} g(x) dx\; .
\eeq
After resummation is achieved, we take the inverse Fourier and Mellin transformations and the result will be convoluted with the two isolated fragmentation functions in~(\ref{taufac}).

\section{Threshold resummation for hadron-pair production toward NNLL \label{sec3}}

\subsection{Resummation formula at next-to-next-to-leading logarithm \label{sec32}}

We start by presenting the resummed cross section and then discuss its structure. The resummed cross section in moment space takes the following form~\cite{KS,BCMN,KOS,KO1,Almeida:2009jt}:
\beeq
\hspace*{-0.7cm}\tilde{\omega}_{ab\to cd}^{\mathrm{resum}}
\left(N,\Delta \eta, \as(\mu_R), \frac{\mu_R}{\hat{m}},\frac{\mu_F}{\hat{m}} 
\right) &=& \xi_R\left(\as(\mu_R), \frac{\mu_R}{\hat{m}} \right)\, \xi_F^{abcd}\left(\as(\mu_R), \frac{\mu_F}{\hat{m}}\right) \nn \\
& \times & \Delta^{N+1}_a \left(\as(\mu_R), \frac{\mu_R}{\hat{m}},\frac{\mu_F}{\hat{m}} \right)
\Delta^{N+1}_b \left(\as(\mu_R), \frac{\mu_R}{\hat{m}},\frac{\mu_F}{\hat{m}} \right) \nn \\[2mm]
&\times& \Delta^{N+2}_c \left(\as(\mu_R), \frac{\mu_R}{\hat{m}},\frac{\mu_F}{\hat{m}} \right)
\Delta^{N+2}_d\left(\as(\mu_R), \frac{\mu_R}{\hat{m}},\frac{\mu_F}{\hat{m}} \right)\nn \\[2mm]
&\times&{\mathrm{Tr}} \left\{ H \left(\Delta\eta,\as(\mu_R) \right)\, {\cal{S}}^\dagger_N 
 \left(\Delta\eta,\as(\mu_R), \frac{\mu_R}{\hat{m}} \right) \,\right.\nn \\[2mm]
&& \left.\hspace*{7.5mm} S\left(\as(\hat{m}/\bar{N}),\Delta\eta \right)  {\cal{S}}_N  \left(\Delta\eta,\as(\mu_R), 
\frac{\mu_R}{\hat{m}} \right) \right\}_{ab \to cd}\,\, .
\label{resumm}
\eeeq
The functions $\xi_{R,F}$ are related to the scale dependence of the resummed cross section but they do not contain threshold logarithms, see~\cite{Hinderer:2014qta} for more details. For every external parton in the hard-scattering, we need to take into account a ``jet function'' $\Delta_i^N$ ($i=a,b,c,d$) which exponentiates logarithms that arise due to soft-collinear gluon emissions~\cite{Almeida:2009jt,Cacciari:2001cw,Sterman:2006hu}. In the $\overline{\rm{MS}}$ scheme, the jet functions are given by~\cite{dyresum,Catani:2003zt,Vogt:2000ci}
\bea
\Delta_i^N\left(\as(\mu_R), \frac{\mu_R}{\hat{m}},\frac{\mu_F}{\hat{m}} \right)
&=& R_i(\as(\mu_R))\, \exp\left\{\int_0^1 dz\,\f{z^{N-1}-1}{1-z}\right.\nn\\[2mm]
&&\hspace*{0cm}\times\, \left[ \int_{\mu_F^2}^{(1-z)^2 \hat{m}^2} \f{d\mu^2}{\mu^2} 
A_i(\as(\mu))+D_i(\as((1-z)\hat m))\right]\Bigg\}\, .
\label{Dfct}
\eea
The functions $A_i$, $D_i$ and $R_i$ may be calculated perturbatively as series in $\as$. The relevant coefficients up to NNLL accuracy can be found in~\cite{dyresum,Hinderer:2014qta,Vogt:2000ci,Catani:2003zt,KT,Moch:2004pa,Catani:2001ic,Harlander:2001is,eric}. The evaluation of the integrals in Eq.~(\ref{Dfct}) up to NNLL accuracy can be found in~\cite{Hinderer:2014qta}. 

In addition, we obtain a trace structure ${\mathrm{Tr}} \{ H {\cal{S}}^\dagger_N S  {\cal{S}}_N\}$ in color space~\cite{KS,KOS} that is associated with large-angle soft emission which is sensitive to the color state of the hard scattering. Each of the factors $H_{ab\to cd}$, ${\cal S}_{N,ab\to cd}$, $S_{ab\to cd}$ 
is a matrix in the space of color exchange operators~\cite{KS,KOS}. The $H_{ab\to cd}$ are the hard-scattering functions and the $S_{ab\to cd}$ are soft functions which may be calculated perturbatively
\ba
H_{ab\to cd}\left(\Delta\eta,\as \right) & = & \left(\f{\as}{\pi} \right)^2\left[H_{ab\to cd}^{(0)}\left(\Delta\eta\right)+
\frac{\alpha_s}{\pi} H_{ab\to cd}^{(1)}\left(\Delta\eta \right)+ {\cal O}(\alpha_s^2)\right] \, ,\nn \\
S_{ab\to cd}\left(\as(\hat{m}/\bar{N}),\Delta\eta \right) & = & S_{ab\to cd}^{(0)}+ \frac{\alpha_s(\hat m/\bar N)}{\pi}
S_{ab\to cd}^{(1)}\left(\Delta\eta\right)+ {\cal O}(\alpha_s^2) \; .
\ea
As an example, we consider the partonic process $qq'\to qq'$ in the next Section. Furthermore, two exponential functions ${\cal S}_{N,ab\to cd}$ appear when solving the renormalization group equation for the soft function
\beeq
\label{GammaSoft}
{\cal S}_{N,ab\to cd} \left(\Delta\eta,\as(\mu_R), \frac{\mu_R}{\hat{m}} 
\right) &=& {\cal P}\exp\left[ 
\frac{1}{2} \int^{\hat{m}^2/\bar{N}^2}_{\hat{m}^2} 
\frac{d\mu^2}{\mu^2} \Gamma_{ab\to cd} 
\left(\Delta\eta,\as(\mu) \right)\right] \, ,
\eeeq
where ${\cal P}$ denotes path ordering. The soft anomalous dimensions $\Gamma_{ab\to cd}$ start at ${\cal O}(\alpha_s)$. The first-order terms $\Gamma^{(1)}_{ab\to cd}$ are presented in~\cite{KS,KOS,KO1,msj}. For resummation at NNLL accuracy, we also need to take into account the second-order contributions $\Gamma^{(2)}_{ab\to cd}$ which were derived in~\cite{twoloopad}.

\subsection{Hard and Soft Matrices}

The leading-order matrices $H^{(0)}_{ab\to cd}$ and $S^{(0)}_{ab\to cd}$ can be found in~\cite{KS,KOS,KO1}. For the $qq'\to qq'$ process, choosing a color octet-singlet basis, they read
\beq\label{H0}
H^{(0)}=
\begin{pmatrix}
\frac{2}{N_{c}^{2}}\frac{s^{2}+u^{2}}{t^2}& 0 \\[2mm] 0 & 0
\end{pmatrix} \, , \quad
S^{(0)}=
\begin{pmatrix}
\frac{N_{c}^{2}-1}{4} & 0 \\[1mm] 0 & N_{c}^2
\end{pmatrix} \, ,
\eeq
where $s,t,u$ are the standard Mandelstam variables. At ${\cal O}(\alpha_s)$, the hard-scattering matrix $H^{(1)}_{ab\to cd}$ can be extracted from purely virtual diagrams. We make use of the 1-loop amplitudes of~\cite{Kunszt:1993sd,Bern:1990cu,Bern:1991aq} which are given in a helicity basis. A similar color decomposition needs to be performed as for the leading-order result. Since the final expression is rather lengthy, we refer the reader to~\cite{Hinderer:2014qta}, where a detailed derivation is presented. Note that a similar calculation within the framework of SCET was carried out in~\cite{Kelley:2010fn,Broggio:2014hoa} and very recently in~\cite{Moult:2015aoa}.
\begin{figure}[t]
\begin{center}
\vspace*{-1cm}
\hspace*{-5.5cm}
\includegraphics[width=8.5cm,angle=90]{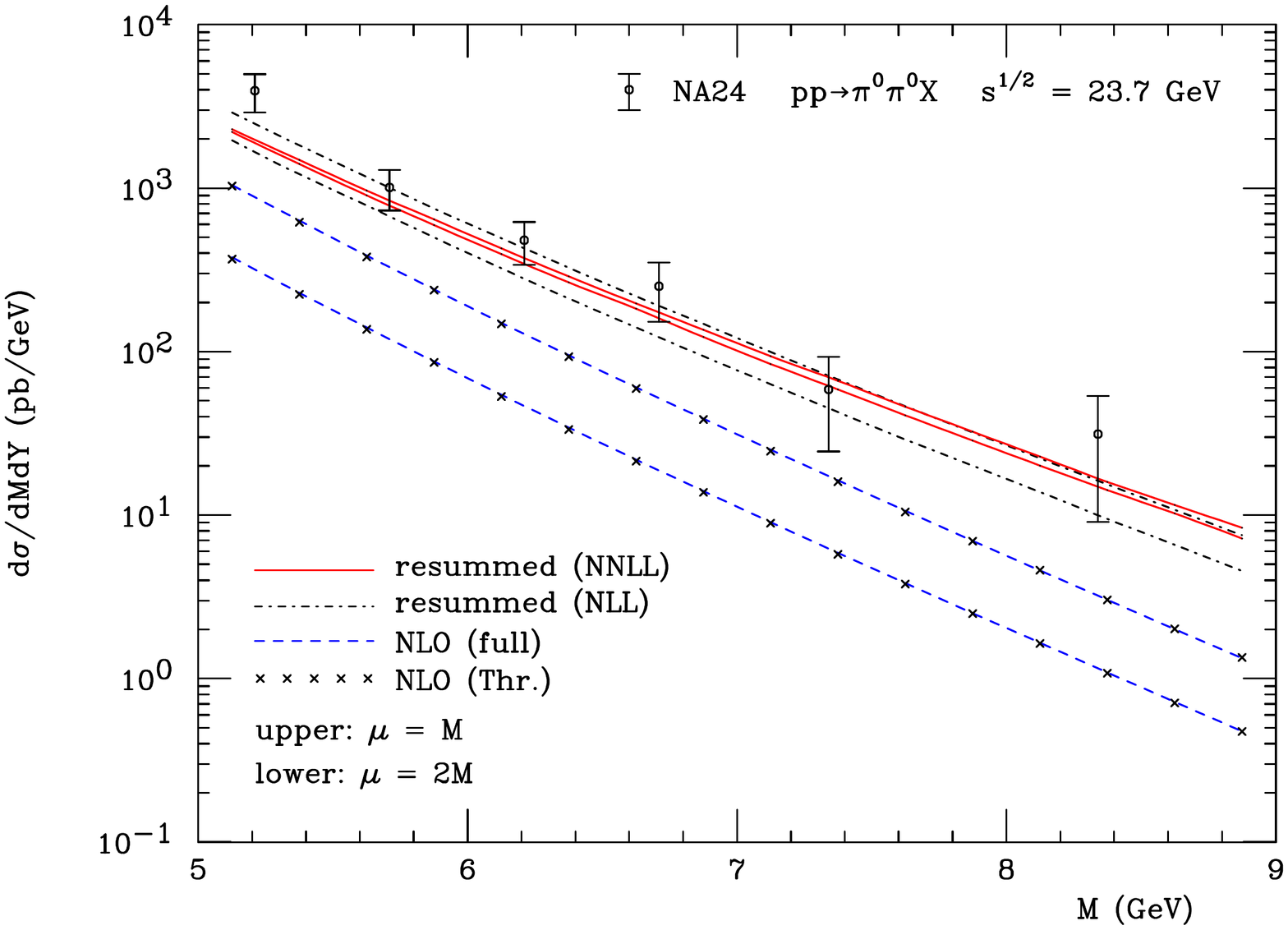}
\hspace*{-1.6cm}
\includegraphics[width=8.5cm,angle=90]{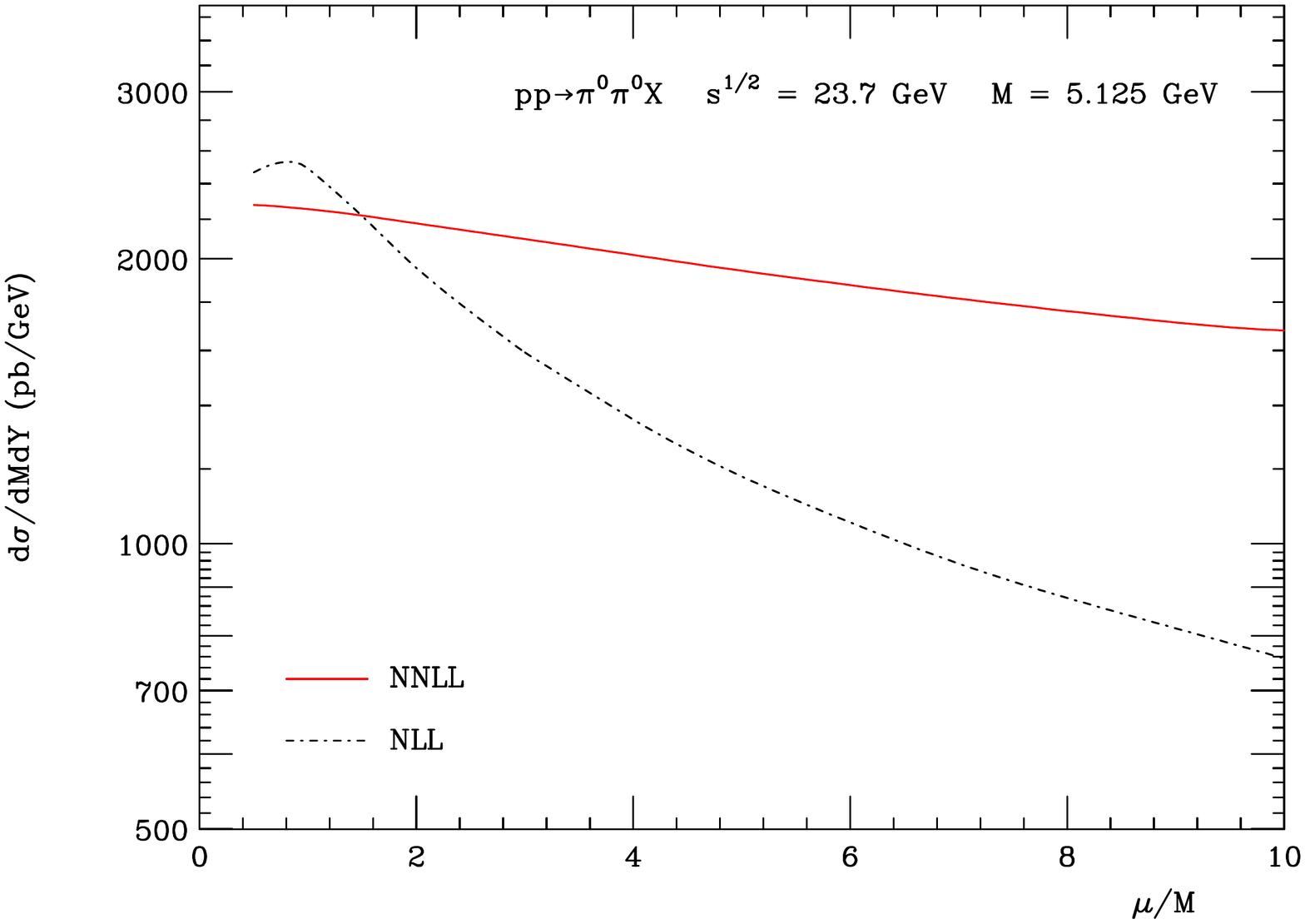}
\hspace*{-5.6cm}
\vspace{-1.5cm}
\end{center}
\caption{\sf \label{fig:1} Di-hadron cross sections for NA24~\cite{na24} kinematics, see text.}
\end{figure}
The soft matrix $S^{(1)}_{ab\to cd}$ at ${\cal O}(\alpha_s)$ can be obtained by analyzing the color structure of $2\to 3$ real-emission diagrams. As an example, we present the explicit result for the $qq'\to qq'$ process
\beq
S^{(1)}\,=\,\frac{C_F}{2}\left(
   \begin{array}{cc}
   {\mathrm{Li}}_2\left(-\frac{u}{t}\right)+
   (2-N_c^2)\,{\mathrm{Li}}_2\left(-\frac{t}{u}\right)&
   -2 N_c\, {\mathrm{Li}}_2\left(-\frac{t}{u}\right)
   \\[2mm]
-2 N_c\, {\mathrm{Li}}_2\left(-\frac{t}{u}\right) &-4 N_c^2\,
 {\mathrm{Li}}_2\left(-\frac{u}{t}\right)
   \end{array}
   \right)\; ,
\eeq
see~\cite{Hinderer:2014qta} for a detailed derivation.

\section{Phenomenological results \label{sec4}}

We present some numerical studies illustrating the effects of threshold resummation at NNLL for di-hadron production. In particular, we compare our new results to the NLL results of~\cite{Almeida:2009jt} and to the full NLO ones of~\cite{Owens:2001rr}. In particular, we compare our new results to the NLL and NLO results of~\cite{Almeida:2009jt}. We compare to the full NLO of~\cite{Owens:2001rr}. Firstly, we consider two examples from~\cite{Almeida:2009jt} concerning the NA24~\cite{na24} (fixed target) and the CCOR~\cite{ccor} (ISR collider) $pp\to\pi^0\pi^0$ scattering experiments. The two experiments were operating at a c.m.s. energy of $\sqrt{S}=23.7$~GeV and $\sqrt{S}=62.4$~GeV respectively. See~\cite{Almeida:2009jt,Hinderer:2014qta} for the experimental cuts employed. For all our numerical calculations presented here, we use the CTEQ6M5 set of parton distribution functions~\cite{cteq6} and the ``de Florian-Sassot-Stratmann'' (DSS) set of
fragmentation functions~\cite{DSS}. We choose the renormalization and factorization scales equal, $\mu_R=\mu_F=\mu$, and always plot the cross section for $\mu=M$ and $\mu=2M$ in order to investigate the QCD scale uncertainty.

On the left side of Fig.~\ref{fig:1}, we show the comparison of NLO (dashed), NLL resummed (dot-dashed) and NNLL resummed (solid) calculations of the di-hadron cross section to the NA24 data. $\mu=M$ corresponds to the upper lines and $\mu=2M$ to the lower lines. The crosses show the NLO expansion of the resummed result which agrees with the full NLO result to a remarkable degree. The NLL resummed cross section has a somewhat steeper slope than the NNLL resummed result. One clearly notices the improved scale dependence when going from NLO to NLL and finally to NNLL accuracy. We study the improved scale uncertainty in more detail on the right side of Fig.~\ref{fig:1}. We show the variation of the cross section as a function of $\mu/M$ for a pair mass of $M=5.125$~GeV which corresponds to the leftmost point on the left side of Fig.~\ref{fig:1}. We plot both the NLL resummed (dot-dashed) and the NNLL resummed (solid) cross sections. As can be seen, the scale dependence at NNLL is almost flat even up to scales as large as $\mu=10M$. 

\begin{figure}[t]
\begin{center}
\vspace*{-1cm}
\hspace*{-5.5cm}
\includegraphics[width=8.5cm,angle=90]{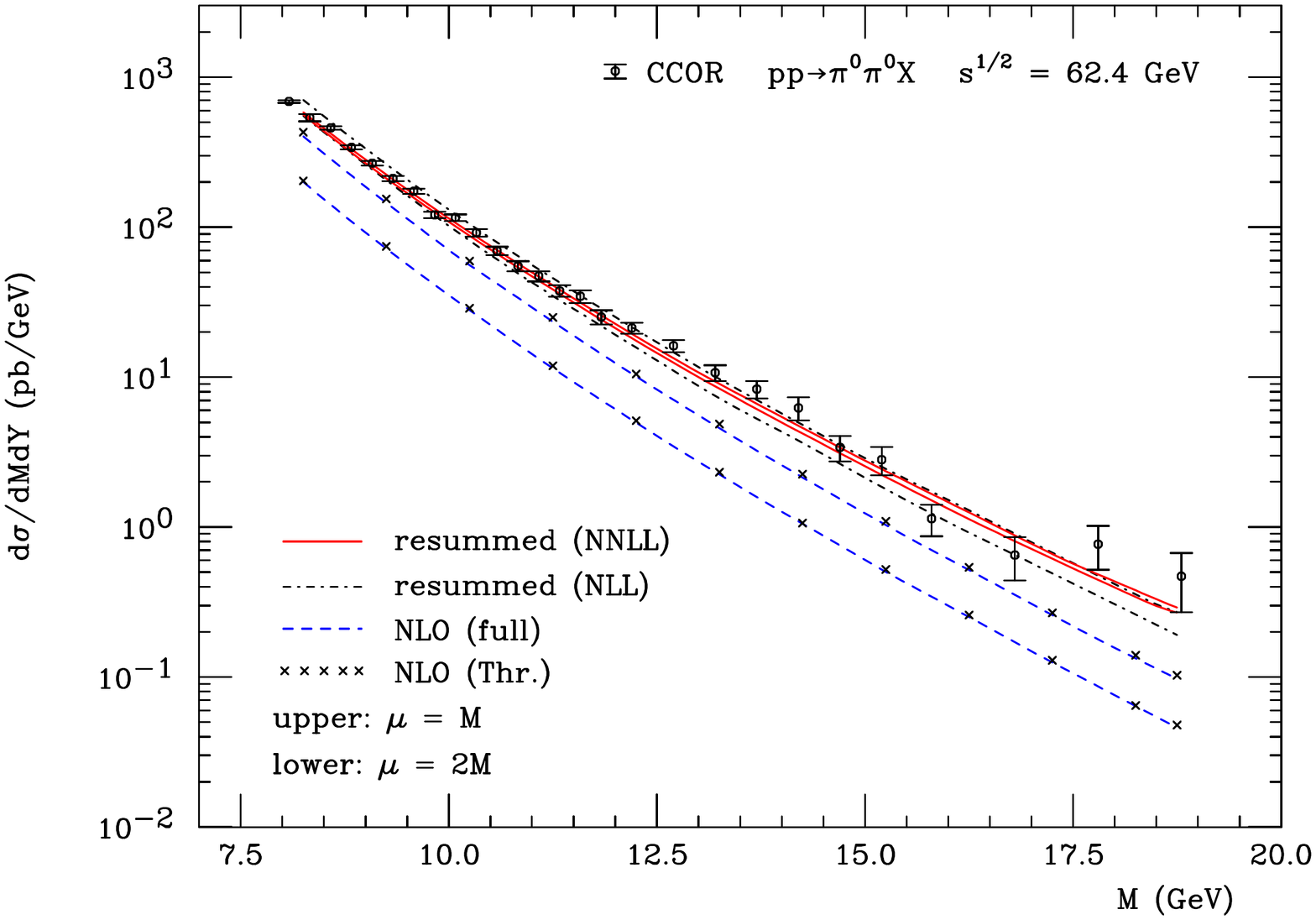}
\hspace*{-1.6cm}
\includegraphics[width=8.5cm,angle=90]{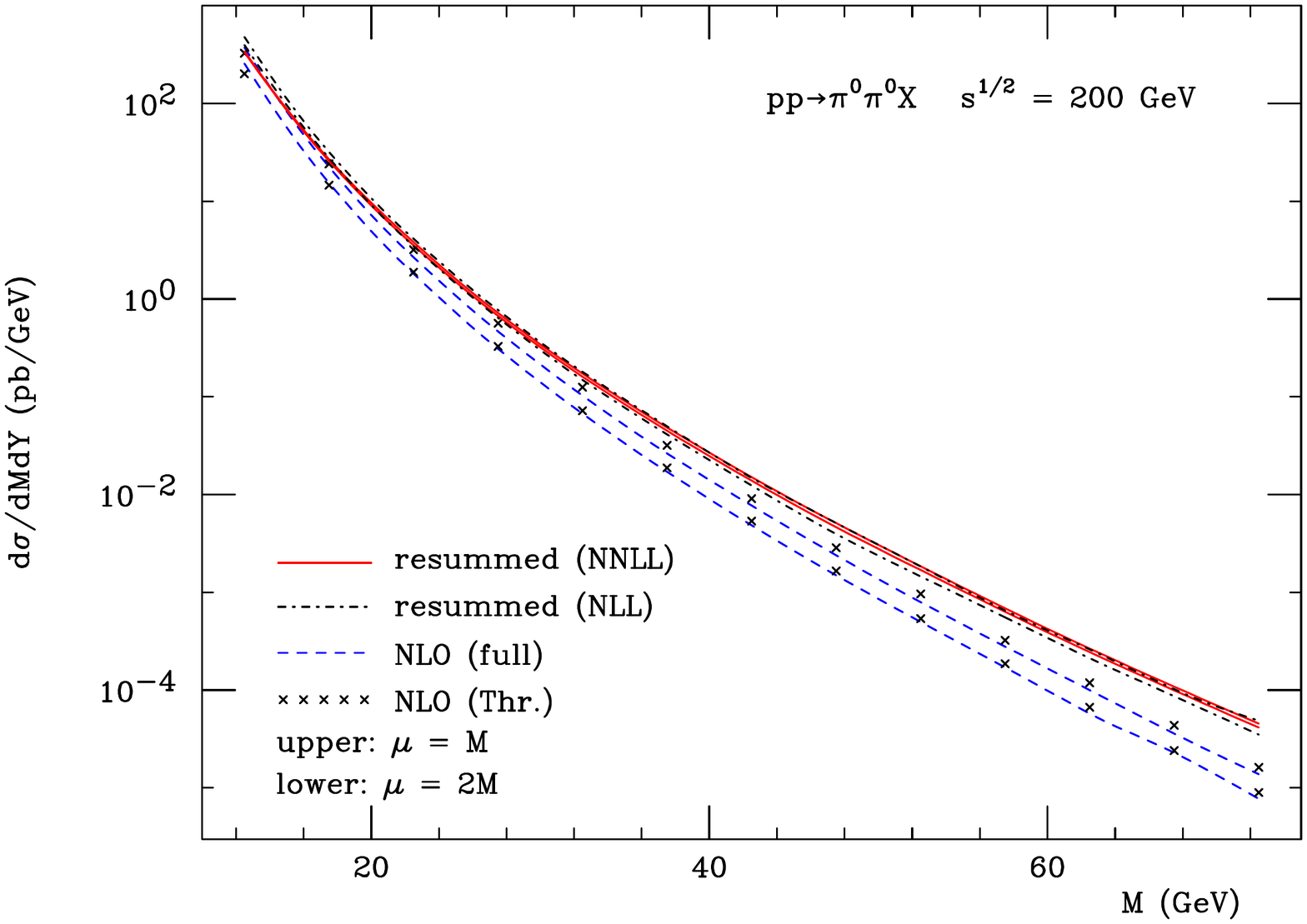}
\hspace*{-5.6cm}
\vspace{-1.5cm}
\end{center}
\caption{\sf \label{fig:2} Di-hadron cross sections for CCOR~\cite{ccor} (left) and RHIC (right) kinematics, see text.}
\end{figure}

Secondly, we compare our theoretical calculations to CCOR data on the left side of Fig.~\ref{fig:2}. The ISR was colliding protons at a c.m.s. energy of $\sqrt{S}=62.4$~GeV. We find very good agreement between our theoretical NNLL calculation and the data. Finally, we also show results for the di-hadron cross section for RHIC energies at $\sqrt{S}=200$~GeV. For both plots in Fig.~\ref{fig:2}, we applied the same kinematical cuts as for the NA24 experiment shown in Fig.~\ref{fig:1}.

\section{Conclusions \label{sec5}}

We have extended the framework of threshold resummation beyond next-to-leading logarithmic accuracy for di-hadron production in hadronic collisions, $H_1 H_2\to h_1 h_2 X$. We have resummed four towers of threshold resummation by taking into account the first-order corrections to the hard-scattering function $H$ and the soft function $S$. Both of these functions are matrices in color space. In our numerical studies, we have found that the scale uncertainty is much reduced compared to previous calculations at NLL or NLO. 

There are important further applications of our work in the context of hadronic collisions. For example, the cross section for single-inclusive hadron production $H_1 H_2\to hX$ is of great phenomenological relevance. The structure of the resummed partonic cross section is in fact similar to the case of di-hadron production~\cite{ddfwv,Catani:2013vaa,deFlorian:2013taa}. With the techniques established in this work, one may also extend the accuracy of resummation for this process toward NNLL. In addition, di-jet and single-inclusive jet~\cite{deFlorian:2007fv} cross sections are of particular interest at present-day collider experiments. 

\section*{Acknowledgments} 

We are grateful to Leandro Almeida, Marco Stratmann, and Ilmo Sung for valuable discussions. 
This work was supported by the ``Bundesministerium f\"{u}r 
Bildung und Forschung'' (BMBF, grant no. 05P12VTCTG). The work of GS was supported in part by the National 
Science Foundation,  grants No. PHY-0969739 and No. PHY-1316617.

\end{document}